\documentstyle[12pt]{article}
\if@twoside  m
\else
\fi
\title{Adiabatic Quantum Transport: Quantization and Fluctuations}
\author{J.~E.~Avron\\ Department of Physics, Technion, 32000 Haifa,
Israel \and R.~Seiler and P.~G.~Zograf \,$^*$\\  Fachbereich Mathematik,
Technische Universit\"at, Berlin 1000, FRG}

\begin{document}
\maketitle
\begin{abstract}
\noindent
Quillen's  local index theorem is used to study the charge transport
coefficients (adiabatic curvature)
associated to the ground state of a Schr\"odinger operator for a
charged (spinless) particle on a closed, multiply
connected surface. The formula splits  the adiabatic curvature
into an explicit integral part and a fluctuating part which has a
natural interpretation in terms of quantum chaos.
\end{abstract}

\vfill

{\it{ \footnotesize ${}^*$  On leave from Steklov Mathematical Institute,
St. Petersburg, Russia}}

\newpage
Some of the interesting recent developments in quantum transport are
connected with two complementary phenomena: the precise quantization
of certain transport coefficients (e.g. conductance) \cite{Hall},
and the fluctuation of others \cite{Altshuler}.
Normally the two are viewed
as disparate phenomena that have little to do with each other: they
pertain to different kinds of physical systems, to different notions
of conductances and the theoretical frameworks used to describe the
two seem to have little in common.

Here we shall study a class of  quantum systems for which the transport
coefficients simultaneously display
quantization and fluctuation, in a way that, we believe, sheds
light on both. A precise description of the kind of systems we
consider shall be given below. For the moment, however, these may be
thought of roughly as describing the dynamics of a Schr\"odinger
particle on a two dimensional multiply connected surface
which has no boundary but is of {\em  finite area}. We consider the
non-dissipative transport in
a constant magnetic field and  $2h$ Aharonov-Bohm flux
tubes that thread the $h$ handles of the surface (for the ordinary
torus $h=1$).

There are two related notions of transport coefficients
that we consider:  {\em conductance} and {\em charge transport}.
Change of flux through the handles generates
electromotive forces around the loops. In the limit of small
emf's the currents and emf's are related by the antisymmetric
conductance matrix. The charge
transport
connects the charges transported around the loops to adiabatic
increase of fluxes
by one quantum unit. They are given by
integrals of the  corresponding elements of the conductance matrix.

The tool we bring to bear is a local index theorem due to Quillen
\cite{Quillen}. When applicable, it splits the conductances
into two parts. The first is explicit and  universal, i.e. it
is - up to factor - the canonical symplectic form on the space of
Aharonov-Bohm fluxes. Its two-dimensional integrals yield quantized
charge transport (in units of $e^2/h$) and therefore provide a
connection to the Integral Quantum Hall effect \cite{Hall, TKNN}.
The second piece in the formula is a complete
derivative, hence it does not affect charge transport. It affects however
the conductance as a fluctuation term. In contrast to the first, it
depends on spectral properties of the Hamiltonian; it is related
to (an appropriate regularization of) its determinant.

In systems which are classically chaotic there is a relation between
the eigenvalues and the lengths of closed classical orbits. This
relation is exact in situations where Selberg's trace formula
applies \cite{Selberg} and is a semi-classical approximation in cases
where Gutzwiller's formula holds \cite{GCB}.  In some special cases
(see below) the determinant of the Hamiltonian can be
expressed in terms of Selberg's zeta function.This connects  conductance
fluctuations  to quantum chaos.

In the general context of adiabatic curvature \cite{Berry}, Quillen's theorem,
 when
applicable, solves a problem posed by Robbins and Berry \cite{RB},
namely, what does the adiabatic curvature say about classically
chaotic systems.

There are several applications to this point of view; all of them are
consequences of formula (15) below:
\begin{enumerate}
\item
All transport coefficients
associated to the ground state for any compact Riemann surface with
h handles  and sufficiently large magnetic field $B$ are explicitly
determined: they are $\pm 1$ if the
pair of fluxes is associated with intersecting loops and $0$
otherwise. This solves a problem posed in \cite{AKPS}, also solved,
by different means,  in \cite{Pnueli}.
\item
The contribution of quantum chaos to conductance fluctuations in the
ground state, in the limit of large magnetic fields, is
exponentially small.  This result has the same flavor as results on
the role of the thermodynamic limit on the quantization of the Hall
conductance \cite{TN}.
\item There are no finite volume correction to
{\em charge transport} in the ground state for quantum systems for
which Quillen's theorem  applies.
\item In special cases where we can apply
Selberg's trace formula, conductance fluctuations in the ground state
are
related to the periodic orbits of the corresponding classical
system.
\end{enumerate}

Now we shall describe the  model in more detail. Consider a charged,
non-relativistic, spinless particle on a topologically closed
surface $\Sigma$ with $h$ handles. Fix a complex structure on
it with local coordinate $z$ and a conformal metric
$\rho (z,\bar z )|dz|^2$. The area element is the 2-form
$\sigma = \frac{i}{2} \rho dz \wedge d\bar z$.

The magnetic field is
a real 2-form $B\sigma$. We allow only constant magnetic
fields (relative to $\sigma$) with total magnetic flux $2\pi\,f= \int_\Sigma
B\sigma$, where $f$ is an integer.

For $B$ fixed consider the space $\cal A$ of real gauge potentials.
For any $A \in \cal A$ we have $ dA = B\sigma$,
where
\begin{eqnarray}
dA = \frac{\partial A}{\partial z} dz +
\frac{\partial A}{\partial {\bar z}} {d \bar z}
\end{eqnarray}
and locally on
$\Sigma$
\begin{eqnarray}
A = {\bar A}^{0,1}(z,\bar z)dz +{A}^{0,1}(z,\bar z)d\bar z.
\end{eqnarray}

To show explicitly the connection between gauge potentials and
Aharonov-Bohm fluxes we introduce on $\Sigma$ a canonical basis
of 1-cycles, i.e. simple oriented loops $\gamma_1,....,\gamma_{2h}$,
which consists of $h$ disjoint pairs $\gamma_j, \gamma_{j+h}$ (one
pair for each handle) such that $\gamma_j$ intersects $\gamma_{j+h}$
exactly at one point. Consider the dual basis $a_1,....,a_{2h}$ of
harmonic one forms on $\Sigma$ normalized by
\begin{eqnarray}
\int_{\gamma_j}a_k =\delta_{jk},\qquad j,k = 1,...2h.
\end{eqnarray}
Fix some $A_0 \in \cal A$; then any gauge potential $A\in\cal A$
is uniquely represented in the form
\begin{eqnarray}
A=A_0 + \sum_{j=1}^{2h} \phi^j a_j + i\,dg\,g^{-1},
\end{eqnarray}
where $g:\Sigma \rightarrow U(1)$ is a gauge transformation.
Factorizing the space $\cal A$ by the action of the gauge group,
we obtain a 2h-dimensional torus $\Phi$ parametrized by the
Aharonov-Bohm fluxes.

The flux torus $\Phi$ carries  natural
symplectic and complex structures.  (Both of them
are necessary to formulate Quillen's theorem).
In terms of the
Aharonov-Bohm fluxes the symplectic structure is given by
\begin{eqnarray}
\Omega = \sum_{i=1}^{h} d{\phi}^j \wedge d{\phi}^{j+h}
\end{eqnarray}.

To define the complex structure, consider the Hodge star operator on
harmonic 1-forms on $\Sigma$,
\begin{eqnarray}
* (\alpha\, dz + \bar \alpha \,d\bar z) =
i (\bar \alpha \,d\bar z - \alpha \,dz)
\end{eqnarray}
Since $*^2 = \, -1$, we can regard $*$ as a multiplication by $i$; this
gives rise to a complex structure on $ \Phi$ which we denote by $J$.
It is explicitly given by
$J=\Omega \,g$, where $g$ is the Riemannian metric on $\Phi$
\begin{eqnarray}
g_{ik} = \int_\Sigma a_i\wedge * a_k.
\end{eqnarray}
Now we return to the discussion of particles on $\Sigma$ in the presence
of a constant magnetic field B. The family of  Schr\"odinger
operators we consider is  defined through the kinetic energy
of a spinless particle on
$\Sigma$ including a gauge potential $A$ and is given by
\begin{eqnarray}
H(\phi) = (d + iA)^* (d + iA)
        = 4 D^*D + B,
\end{eqnarray}
where $D = \frac{\partial}{\partial \bar z} + i A^{0,1}$ and
$D^* = - \frac{1}{\rho}(\frac{\partial}{\partial z} + i \bar A^{0,1})$. The
Schr\"odinger operator acts on the Hilbert space $\it H$ of sections
of a U(1) line bundle with the scalar product
\begin{eqnarray}
(\psi,\psi) = \int_\Sigma |\psi|^2\,\sigma.
\end{eqnarray}

Let us recall a few basic facts about adiabatic transport \cite{Hall,TKNN}.
Let $P(\phi)$ denote the spectral projection on the ground state of
$H(\phi)$ \cite {zero energy}. Suppose that the fluxes $\phi^j(t)$ depend
 adiabatically
on time. Then,
the current around the k-th flux is given by
\begin{eqnarray}
I_{k}(P,\phi) =
i \sum_{j=1}^{2h}\dot{\phi}^j\,\omega_{jk}(P,\phi).
\end{eqnarray}
Here $\omega_{jk}(P,\phi)$ is the $jk$ component of the
adiabatic curvature 2-form $\omega\,(P,\phi)$ \cite{Berry}:
\begin{eqnarray}
\omega(P,\phi)= {Tr\,}(P {\bf d}P\wedge {\bf d}P)={
Tr} \left(P ({\bf d} H)\, \widehat{(H-E)}^{-2} ({\bf d}H)
\right),
\end{eqnarray}
where $\bf d$ denotes exterior derivative with respect to flux
and ${\bf d }P = \sum \frac{\partial P}{\partial \phi^i} {\bf d}\phi^i$.
The hat in $ \widehat{(H-E)}^{-1}$ excises the pole at $E$ (the
reduced resolvent). The expression on the
right hand side of Eq.\,(11) is Kubo's formula in operator notation. In
the case at hand, the adiabatic curvature is also given by the formula
\begin{eqnarray}
\omega(P,\phi)= {Tr\,}(P {\bf d}{D^*}\frac{1}{DD^*} {\bf d}D).
\end{eqnarray}

The periods of the current one form have a direct physical interpretation.
The total charge transported around the k-th
flux driven by the adiabatic increase by one unit of
quantum flux  in the $j$-th flux is:
\begin{eqnarray}
Q_{jk} = \int_{0\le \phi^j\le 2\pi}\omega_{jk}(P,\phi) d\phi^j.
\end{eqnarray}

In the situation we consider Quillen's local index theorem
gives an explicit formula for the curvature form of the so-called
determinant bundle of the family of Cauchy-Riemann operators $D$,
parametrised by Aharonov-Bohm fluxes $\phi$. To apply Quillen's formula
to the analysis of adiabatic curvature, notice first that by Riemann-Roch
\begin{eqnarray}
dim\,ker\,D - dim \,ker\, D^* = f - h + 1.
\end{eqnarray}
By a standard positivity argument $ker \, D^*  = 0$ when
$f\ge 2h -1$. Then $dim\, ker\, D =  f - h + 1$
is constant on the flux space and
 Quillen's theorem yields the following formula for conductance:
\begin{eqnarray}
\frac{i}{2\pi}{\omega} (P,\phi) = - {\frac{1}{(2\pi)^2}} \Omega -
{\frac{1}{4\pi}}\,d\,J\,d \,{\rm log \,det}\, D^{*}D.
\end{eqnarray}
Here  ${\rm det} \,D^{*}D$ is
defined via the zeta function regularization,
\begin{eqnarray}
\log\,\det \,D^{*}D =\lim_{\varepsilon \to 0}\, \int
_\varepsilon^\infty {\rm Tr}\left(e^{-t \,D^{*} D }\right)
\frac{dt}{t}.
\end{eqnarray}
The
determinant term is an exact two form on
$\Phi$, hence it does not contribute to charge transport,
and
represents conductance fluctuations.
The natural symplectic form $\Omega$ on
the flux torus depends only on the topology of the
surface  and in this sense is universal. Furthermore it
determines the charge transport. Thus, Quillen's
theorem divides the conductance into a constant and a fluctuating
term.

In addition to the consequences of formula (15) mentioned at the
beginning we can draw the following conclusions:
\begin{enumerate}
\item Formula (15) suggests, that the behavior of conductance is qualitatively
more complicated for small magnetic fields,
i.e. $f = 0,1, ....,2(h-1)$.
\item It is known that for the special case of the flat two torus
${\rm det}\, D^{*}D$ is constant on the flux
torus. Hence conductance has no fluctuating  term. This relates well
to the fact that classical dynamics on the flat torus is integrable.
\item More interesting is the case of surfaces of
constant negative curvature $-1$. They are conveniently
realized as the orbit spaces
of  discrete subgroups $\Gamma$ of $SL(2,R)$ acting on
the upper half plane $H$ . In this case the determinant of
the Hamiltonian can be expressed
in terms of Selberg's zeta function,
\end{enumerate}
\begin{eqnarray}
Z(s,\phi) = \prod_{primitive\, \gamma}
\prod_{k=0}^\infty \left(1-\chi(\gamma,\phi) e^{-(s+k)\ell(\gamma)}
\right).\end{eqnarray}
Here $\gamma \in \Gamma$ are primitive hyperbolic elements of $\Gamma$
representing conjugacy classes and  may be
thought of as closed geodesics on $\Sigma$;  $\ell(\gamma)$ denotes the length
of $\gamma$ and  $\chi(\gamma,\phi)$ are
phase factors that carry information about the fluxes:
\begin{eqnarray}
\chi(\gamma,\phi) =
 \exp i\,\left(\sum_1^{2h} \phi_j \,
n_j(\gamma)\right)
\end{eqnarray}
where  $n_j(\gamma)$ counts the
number of times (including signs) the closed geodesic $\gamma$  goes
around the j-th fundamental loop.
The determinant of $ D^{*} D$ is expressed in terms of Selberg zeta function
by \cite{DP}:
\begin{eqnarray}
\det D^{*} D = c_h \, Z(B,\chi) ,
\end{eqnarray}
$c_h$ is a constant independent
of the fluxes.
When B becomes large,
$Z(B,\chi)$ tends  to one and its derivatives decrease exponentially.
Therefore the chaotic contribution to conductance is
exponentially small for large magnetic fields. Moreover the leading
term in the asymptotics of the fluctuating term for large B is
given by $ \exp -\left(B\ell(\gamma_{min})\right)$,
where $\gamma_{min}$ is the shortest homologically non trivial
closed geodesic.

\section*{Acknowledgment}
This work is supported in part by the DFG, the GIF and by the Fund
for the Promotion of Research at the Technion. RS and PZ acknowledge
the hospitality of the ITP at the Technion.


\begin{thebibliography}{article}
\bibitem{Hall}  See e.g.\, R.E.~Prange and
S.M.~Girvin, {\em ``The Quantum Hall Effect"},Springer (1987);
M.~Stone, Ed.  {\em ``Quantum Hall Effect"}, World Scientific,
Singapore (1992).
 \vspace{-1.8ex}
\bibitem{Altshuler}
B.L.~Altshuler and B.I. Shklovskii, Zh.~Exp.~Teor.~Fiz. {\bf 91},220
(1986), [Sov.~Phys.~JETP {\bf 64}, 127, (1986)]; B.L.~Altshuler, in
{\em Proceedings of the International Symposium on Nanostructure and
Mesoscopic Systems}, W.P.~Kirk Ed. Santa Fe (1991); N. Argaman, Y.
Imry and U. Smilansky, Phys. Rev. {\bf B 47}, 4440 (1993); E.~Doron
and U.~Smilansky, Phys.~Rev.~Lett. {\bf 65}, 3072, (1990); R.A.
Jalabert, H.U. Baranger and A.D. Stone, Phys. Rev. Lett. {\bf 65},
2442 (1990); E. Akkermans, Physica A, {\bf 200}, 530-537 (1993).
   \vspace{-1.8ex}
\bibitem{Quillen}
D. Quillen, Funk.~Anal.~Prilozh. {\bf 19}, 37 (1985)
[Funct.~Anal.~{\bf 19} 31-34 (1985)]; L. Alvarez-Gaum\`{e},
G.~Mourre and C.~Vafa, Comm.~Math.~Phys. {\bf 106}, 1-40,(1986).
   \vspace{-1.8ex}
 \bibitem{TKNN}  D.~J.~Thouless, M.~Kohmoto,~P.~Nightingale and M.
den Nijs, Phys.~Rev.~Lett. {\bf 49}, 40, (1982); J.E.~Avron,
R.~Seiler and L.~Yaffe, Comm.~Math.~Phys., {\bf 110}, 33,
(1987).
    \vspace{-1.8ex}
\bibitem{Selberg}
D.A.~Hejhal {\em The Selberg trace formula for PSL(2,R)} vol.\,I,
Springer LN in Math. {\bf 548}, Springer, (1976).
   \vspace{-1.8ex}
\bibitem{GCB} M.C.~Gutzwiller, in {\em ``The Selberg Trace formula"},
D.~Hejhal, P.~Sarnak and A.~Terras Eds. Contemporary Math. {\bf 53},
AMS, Providence, (1984); N.~Balasz and A.~Voros {\em Chaos and the
Pesudosphere}, Phys.~Rep. {\bf 143} 103, (1986); See also {\em Proceeding of
the
Houches summer school}, Eds. M.J. Gianoni and A. Voros, North
Holland, Amsterdam (1989); M.C.~Gutzwiller,
{\em Chaos in Classical and Quantum Mechanics}, Springer, (1990);
M.~Antoine, A.~Comtet and S.~Ouvry, J.Phys. A. {\bf 23} 3699,
(1990); E.B.~Bogomolny, B.~Georgeot, M.J.~Giannoni and C.~Schmidt,
Phys.~Rev.~Lett. {\bf 69}, 1477-1480, (1992);
 \vspace{-1.8ex}
 \bibitem{Berry}
M.V.~Berry, Proc.~Roy.~Soc.~A {\bf 392},45-57,(1984); B.Simon,
Phys.~Rev.~Lett. {\bf 51} 2167-2170 (1983); A.A. Shapere and F.
Wilczek,{\em Geometric Phases in Physics}, World Scientific, (1989).
   \vspace{-1.8ex}
\bibitem{RB}
J.~Robbins and M.V.~Berry, Proc.~Roy.~Soc.~Lond A{\bf 436} 631-661,
(1992).
   \vspace{-1.8ex}
\bibitem{AKPS}
J.~E.~Avron, M.~Klein,  A.~Pnueli and L.~Sadun,
Phys.~Rev.~Lett.~{\bf 69}, 128-131, (1992).
   \vspace{-1.8ex}
\bibitem{Pnueli}
A. Pnueli, Private communication.
   \vspace{-1.8ex}
\bibitem{TN}
Q.~Niu and D.J.~Thouless  , Phys.~Rev. B {\bf 35}, 2188-2197,(1987).
\vspace{-1.8ex}
\bibitem{zero energy}
In the situations we study the ground state energy vanishes identically.
\vspace{-1.8ex}
\bibitem{DP}
E.~D'Hoker and D.H.~Phong, Nucl.~Phys.~{\bf B 269} 205-234 (1986);
Rev.~Mod.~Phys.~{\bf 60} 917-1054, (1988);  P.~Sarnak in {``\em
Number Theory, Trace formula and Discrete groups''}, K.E.~Aubert et.
al. eds. Academic Press (1989).
 \vspace{-1.8ex}
\bibitem{Rem2} It turns out
that the lengths of closed geodesics equals the action associated to
the classical orbit in magnetic field.
This is discussed in A.~Comtet, B.~Georgeot and S.~Ouvry,
Phys.~Rev.~Lett.~{\bf 71}, 3786, (1993).
\end{thebibliography}
\end{document}